\documentclass[aps,prb,reprint]{revtex4-2}

\usepackage{cancel}
\usepackage{graphicx}
\usepackage{bm}
\usepackage{bbold}
\usepackage[breaklinks=true, colorlinks, citecolor=blue, urlcolor=blue]{hyperref}
\usepackage{tikz}
\usepackage[compat=1.1.0]{tikz-feynman}
\usepackage{amsmath}
\usepackage{csquotes}

\begin{document}

\def\SigmaDiagram{\begin{tikzpicture}[baseline=(a.base)]
\begin{feynman}
\vertex(a) at (0,0);
\vertex(b) at (1.5,0);
\vertex(m) at (0.75,0){};
\vertex[below = 0.4 cm of m](m1){\(\mathbf{k}^{\prime},i\omega_{n}^{\prime}\)};
\vertex[above = 0.8 cm of m](m2){\(V(\mathbf{k}-\mathbf{k}^{\prime})\)};
\diagram*{
(a)--[fermion](b),
(a)--[scalar, out=70,  in=110, looseness=1.2](b),
};
\end{feynman}
\end{tikzpicture}
}

\newcommand{\be}{\begin{eqnarray}}
\newcommand{\ee}{\end{eqnarray}}
\newcommand{\beq}{\begin{equation}}
\newcommand{\eeq}{\end{equation}}
\newcommand{\nn}{\nonumber\\}
\newcommand{\pr}{^{\prime}}
\newcommand{\ph}{^{\text{ph}}}
\newcommand{\ct}{^{\text{CT}}}
\newcommand{\vf}{v_{\text{F}}}
\newcommand{\FD}{\mathcal{F}}
\newcommand{\vare}{\varepsilon}
\newcommand{\nf}{n_{\text{F}}}
\newcommand{\fe}{f_{e}}
\newcommand{\fh}{f_{h}}
\newcommand{\kb}{k_{\text{B}}}
\newcommand{\unit}{\sigma_{0}}
\newcommand{\tr}{\operatorname{Tr}}
\renewcommand{\Re}{\operatorname{\mathfrak{Re}}}
\renewcommand{\Im}{\operatorname{\mathfrak{Im}}}
\newcommand{\blue}[1]{\textcolor{blue}{#1}}

\preprint{APS/123-QED}

\title{First-order effect of electron-electron interactions on the anomalous Hall conductivity of massive Dirac fermions}

\author{A. Daria Dumitriu-I.}
\email{alexandra-daria.dumitriu-iovanescu@outlook.com}
\affiliation{Department of Physics and Astronomy, University of Manchester, Manchester M13 9PL, United Kingdom}
\author{Darius-A. Deaconu}%
\affiliation{Department of Physics and Astronomy, University of Manchester, Manchester M13 9PL, United Kingdom}%
\author{Alexander E. Kazantsev}%
\affiliation{Department of Physics and Astronomy, University of Manchester, Manchester M13 9PL, United Kingdom}%
\author{Alessandro Principi}%
\email{alessandro.principi@manchester.ac.uk}
\affiliation{Department of Physics and Astronomy, University of Manchester, Manchester M13 9PL, United Kingdom}%

\date{\today}

\begin{abstract}
We investigate the first-order correction to the anomalous Hall conductivity of 2D massive Dirac fermions arising from electron-electron interactions. In a fully gapped system in the limit of zero temperature, we find that this correction vanishes, confirming the absence of perturbative corrections to the topological Hall conductivity. At finite temperature or chemical potential, we find that the total Hall response decays faster than in the noninteracting case, depending on the strength of electron-electron interactions. These features, which could potentially be observed experimentally, show the importance of two-body interactions for anomalous Hall transport.

\end{abstract}


\maketitle

\section{Introduction}

After the discovery of the quantum Hall effect, theoretical effort was directed towards understanding the robustness of the quantization of the Hall conductivity $\sigma_H$, which in the absence of interactions is related to a topological invariant~\citep{thouless1982quantized,xiao2010berry}. In a seminal paper~\cite{coleman1985no}, Coleman and Hill showed that two-particle interactions do not modify the value of $\sigma_H$ at zero temperature, if the Fermi energy lies in the bulk band gap, which was later generalized to nonrelativistic interactions in Ref.~\cite{zhang2020influenceanomalous}. This raises the question whether the Hall conductivity is robust to the effects of interactions in systems at finite temperature or chemical potential. To answer this question, we investigate the impact of electron-electron (e-e) interactions in the archetypal model of the anomalous Hall effect~\cite{haldane1988model}, a two-dimensional system of massive Dirac fermions. 

Aside from its foundational significance, this issue also has practical implications since real-world materials exist at nonzero temperatures and are rarely completely free of doping. It has been demonstrated that certain many-body interactions, for example, that between electrons~\cite{lai2014effects,hankiewicz2006coulomb,lee2011effects}, or that between electrons and phonons~\cite{zhang2024quantum}, can substantially influence Hall responses. For instance, in the anomalous Hall and spin Hall effects, the presence of quenched disorder can yield outcomes remarkably at odds with noninteracting results ~\cite{ado2017sensitivity,ado2015anomalous}. These include a faster decay of Hall responses with growing chemical potential, changes in sign, and, in some cases, even the complete elimination of these effects. However, while it may be possible to mitigate disorder in a material by refining the growth process, the omnipresent e-e interactions cannot be easily eliminated. Thus, it is fundamental to understand how they affect the Hall response to be able to predict and explain experimental results.

In this paper, we study the correction to the anomalous Hall effect of massive Dirac fermions to first order in the strength of e-e interactions. We start by introducing the model Hamiltonian and calculate the zeroth- and first-order response functions. We find that the first-order response is divergent. Divergences can be removed by renormalizing the bare parameters, which enter the model. In this way, we obtain a finite expression for the first-order correction to the Hall conductivity.

The first-order correction naturally depends on the form of the e-e interaction. Here, we initially address the case of a contact potential, whose relative simplicity allows us to do significant analytical progress. We relate such a potential to the Coulomb interaction between electrons in an overscreened regime. We show that  interaction corrections can be large enough to counterbalance the noninteracting contribution to the Hall conductivity. In particular, we show that the anomalous Hall conductivity can decay faster with chemical potential than what is predicted in the noninteracting case. We then compute the same correction for an unscreened Coulomb potential, and show that it exhibits similar features. Thus, we confirm that they are robust irrespective of the precise form of the interaction.

\section{Description of the model}

We consider massive Dirac fermions in two dimensions, whose dynamics is described by the many-body Hamiltonian (hereafter, we set $\hbar = 1$)
\begin{align}
\hat{\mathcal{H}}=&\sum_{\alpha\beta,\bm k}\hat{\psi}^{\dagger}_{\alpha,\bm k}(\bm d(\bm k) \cdot \bm \sigma_{\alpha\beta}-\mu \,\delta_{\alpha\beta}) \hat{\psi}_{\beta,\bm k}\nn
&+\frac{1}{2}\sum_{\bm q\neq0}\sum_{\alpha\bm k,\beta\bm k\pr}\hat{\psi}^{\dagger}_{\alpha,\bm k-\frac{\bm q}{2}}\hat{\psi}^{\dagger}_{\beta,\bm k\pr+\frac{\bm q}{2}}V_{\bm q}\hat{\psi}_{\beta,\bm k\pr-\frac{\bm q}{2}}\hat{\psi}_{\alpha,\bm k +\frac{\bm q}{2}},
\end{align}
where $\psi_{\lambda,\bm k}$ $(\psi^{\dagger}_{\lambda,\bm k})$ destroys (creates) a particle with momentum $\bm k$ and pseudospin $\lambda$, $\mu$ is the chemical potential, $H(\bm k)=\bm d(\bm k) \cdot \bm \sigma$ is the single-particle Hamiltonian, $\bm d(\bm k)=(\vf k_x, \vf k_y, \Delta)$ and $\bm \sigma=(\sigma_x, \sigma_y, \sigma_z)$ is a vector of Pauli matrices. Here, $\Delta$ is half the band gap, while $\vf$ is the Fermi velocity. The energy of these particles is $\pm \vare_{k}$,  where $\vare_{k}=\sqrt{(\vf k)^2+\Delta^{2}}$ and the $\pm$ sign identifies the conduction and valence bands, respectively. Finally, $V_{\bm q}$ is the e-e interaction, which will be specified later. The noninteracting Matsubara Green's function (MGF) corresponding to our Hamiltonian takes the form $G^{(0)}(\bm k, i\omega_{n})=\big[(i\omega_{n}+\mu)\unit-H(\bm k)\big]^{-1}$, where $i\omega_{n}$ is a fermionic Matsubara frequency and $\sigma_{0}$ is the unit matrix.

Within the linear-response formalism~\cite{vignale2005quantum}, the Hall conductivity is defined in terms of the current-current response function $\chi_{j_{x}j_{y}}(\bm q, \omega)$ as
\be
\label{eq:kubo}
\sigma_{H}
=\lim_{\omega\to0}\left[ \frac{ie^2}{\omega} \chi_{j_{x}j_{y}}(\bm q=0, \omega)\right]\!.
\ee

\begin{figure}[t]
    \includegraphics{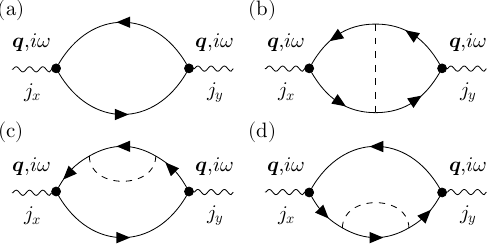}
    \caption{(a) Noninteracting contribution to the current-current response function $\chi_{j_{x}j_{y}}(\bm q, i\omega)$. [(b)--(d)] The first-order contributions to $\chi_{j_{x}j_{y}}(\bm q, i\omega)$ due to vertex corrections (b), and self-energy insertions [(c)--(d)]. In these diagrams, solid lines are MGFs, dashed lines represent electron-electron interactions, while solid dots stand for current vertices.
    }
    \label{fig:1}
\end{figure}

Figure~\ref{fig:1} summarizes the diagrammatic calculation of $\chi_{j_{x}j_{y}}(\bm q, i\omega)$ to zeroth [panel (a)] and first order [panels (b)--(d)] in the e-e interaction. There, solid lines represent noninteracting MGFs, while dashed lines represent the e-e interaction. Finally, the solid dots are the current operators $j_{x/y}=\vf \sigma_{x/y}$.

\section{Zeroth-order contribution}

At zeroth order in interaction we find the Hall conductivity~\cite{vignale2005quantum}(see Appendix~\ref{appendix_A})
\be
\label{eq:sigma_H_0}
\sigma_{H}^{(0)}(\mu,T)=-\frac{e^{2}}{2h}(1-\FD_{-2}^{+}(\bar{\mu},\bar{T})),
\ee
which recovers~\footnote{Since our calculation pertains to a single valley, it only recovers {\it half of} the result for the Haldane model.} the well-known zero-temperature result for the anomalous Hall conductivity at a finite doping (see, e.g., Ref.~\cite{sinitsyn2006charge}), as well as the finite-temperature generalization of the Chern number, the so-called \enquote{Ulhmann number}~\cite{leonforte2019haldane}.
In Eq.~(\ref{eq:sigma_H_0}) we defined the integral 
\be
\label{eq:fermi_dirac_integral}
\FD_{n}^{\pm}=\!\int_{1}^{\infty} \!\!\!dx ~x^{n} \!\!\left( \frac{1}{e^{(x-\bar{\mu})/\bar{T}}+1}\!\pm\! \frac{1}{e^{(x+\bar{\mu})/\bar{T}}+1}\right)\!\!,
\ee
where $\bar{\mu}=\mu/\Delta$, $\bar{T}=\kb T/\Delta$ are the chemical potential and the temperature, measured in units of (half) the gap.\\

\section{First-order contribution}

Figures~\ref{fig:1}(b) and~\ref{fig:1}(c)--\ref{fig:1}(d) show the first-order diagrams, which yield the exchange and self-energy corrections to $\sigma_H$, respectively~\footnote{Note that there are 3 diagrams which we omitted in the main text, namely the RPA and the Hartree self-energy diagrams. The first only contributes to the screening of the external field, while the latter two are canceled by the uniform positive background.}.

We start by evaluating the self-energy insertion, i.e.
\be
\label{eq:sigma_diagram}
-\Sigma(\bm k) = \SigmaDiagram = -\Sigma_{0}(\bm k) \unit - \bm \Sigma(\bm k) \cdot \bm \sigma,
\ee
where
\begin{gather}
\label{eq:sigma_0}
\Sigma_{0}(\bm k)\, =-\frac{1}{2} \int \frac{d^2{\bm k\pr}}{(2\pi)^{2}}\,
V(\bm k-\bm k\pr)\Phi^{+}_{k\pr},\\
\label{eq:sigma_bold}
\bm \Sigma(\bm k)\, =\frac{1}{2} \int \frac{d^2{\bm k\pr}}{(2\pi)^{2}}\,
\frac{\bm d(\bm k\pr)}{\vare_{k\pr}}V(\bm k-\bm k\pr)\Phi^{-}_{k\pr},
\end{gather}
are independent of the external frequency. We defined \(\Phi^{\pm}_{k}
=1-\fh(\vare_{k})\pm \fe(\vare_{k})\), where \(\fe(\vare)\!=\![e^{( \vare-\mu)/(\kb T)}\!+\!1]^{-1}\) and \(\fh(\vare)\!=\![e^{(\vare+\mu)/(\kb T)}\!+\!1]^{-1}\) are the Fermi--Dirac distribution functions for electrons and holes, respectively.

Adding the contributions from diagrams \ref{fig:1}(b)-- \ref{fig:1}(d), and expanding to first order in frequency, we find the first-order correction to the current-current response function
\begin{widetext}
\be
\label{eq:all_diagrams}
\chi_{j_{x}j_{y}}^{(1)}(\bm q=0,i\omega) \to &&- \frac{\omega\vf^2\Delta}{4}
\int \frac{d^2{\bm k}}{(2\pi)^{2}}\,
\Bigg\{-\Bigg[\frac{2\Sigma_{0}(\bm k)}{\vare_{k}^{3}}\frac{\partial \Phi^{+}_{k}}{\partial\vare_{k}}\Bigg]+\Bigg[\frac{2 \Sigma_{\perp}(\bm k)}{\vare_{k}^{2}\Delta}\frac{\partial \Phi^{-}_{k}}{\partial\vare_{k}}\Bigg]\nn
&& +
\int \frac{d^2{\bm k\pr}}{(2\pi)^{2}}\,
 V(\bm k-\bm k\pr)\frac{\vf^2(\bm k \cdot \bm k\pr- k^{\prime 2})}{\vare_{k}^{3}\vare_{k\pr}^{2}}\frac{\partial \Phi^{-}_{k\pr}}{\partial \vare_{k\pr}}\Phi^{-}_{k}\Bigg\},
\ee
\end{widetext}
where we split $\bm \Sigma$ into $\bm \Sigma_{\parallel}=(\Sigma_{x},\Sigma_{y})$ and $\Sigma_{\perp}= \Sigma_{z}$. In~Appendix~\ref{appendix_B} we show that the terms in the self-energy containing $\bm  \Sigma_{\parallel}$ cancel similar terms from the exchange diagram. This is a consequence of the Ward identity~\cite{peskin2018introduction}.

Although the contribution from $\bm \Sigma_{\parallel}$ cancels between diagrams, Eq.~\eqref{eq:all_diagrams} still depends on $\Sigma_{0}(\bm k)$ and $\Sigma_{\perp}(\bm k)$, both of which suffer from ultraviolet (UV) divergences and must be regularized by introducing a cutoff $\Lambda$ (see below). For a contact potential $V({\bm k}) = V_{c}$, where $V_c$ is a constant, which does not depend on the momentum carried by the interaction, these divergences are quadratic $\propto \Lambda^2$ (for $\Sigma_0$) or linear $\propto \Lambda$ (for $\bm \Sigma_{\parallel}$ and $\Sigma_{\perp}$). For a Coulomb potential $V({\bm k})=2\pi e^2/(\kappa k)$ for a medium with dielectric constant $\kappa$, the quadratic divergence becomes linear, while the linear ones become logarithmic.

\subsection{Regularization of divergences}

In order to deal with these divergences, we introduce the UV cutoff in momentum $\Lambda$ by employing a type of Pauli--Villars~\cite{peskin2018introduction}\footnote{We note that dimensional regularization could be used as well. In the case of a contact potential none of the divergences would appear, as they are all positive powers of the cutoff.} regularization scheme: The interaction is modified in such a way as to vanish fast enough at large momentum, so that the integrals become convergent. More explicitly, the interaction becomes \(V(\bm k) \to V(\bm k)\big[1+(|\bm k|/\Lambda)^{n}\big]^{-1}\), where $n$ is a natural number. 
The integrals in Eq.~\eqref{eq:all_diagrams} then contain cutoff-dependent pieces signifying the presence of divergences, as well as finite parts. The latter become independent of the cutoff in the limit $\Lambda\to \infty$. 

UV divergences are absorbed into the redefinition of the parameters, which enter into the Hamiltonian. Bare parameters ($x^{\text{bare}}$, where $x\equiv \mu, \vf, \Delta$) are defined in terms of physical ones ($x^{\text{ph}}$) as $x^{\text{bare}}= x^{\text{ph}}+\delta x\ct$. Here, $\delta x\ct$ are the counterterms. From here on, we assume the Hamiltonian is defined in terms of renormalized parameters and hence omit the label \enquote{ph}.

Using the Dyson equation and the expression given in Eq.~\eqref{eq:sigma_diagram}, we write the dressed MGF as a function of the bare parameters
\begin{align}
\label{eq:dressed_green}
G^{-1}(\bm k)=&\left(i\omega_{n}+\mu+\delta \mu\ct-\Sigma_{0}(\bm k)\right)\unit\nn
&-\left(\bm d(\bm k) + \delta\bm d\ct(\bm k)+\bm \Sigma(\bm k)\right)\cdot \bm \sigma,
\end{align}
where $\delta\bm d\ct(\bm k)=(\delta\vf\ct k_{x},\delta\vf\ct k_{y},\delta\Delta\ct)$. Thus, the divergences introduced by $\Sigma_{0}(\bm k)$, $\bm \Sigma_{\parallel}(\bm k)$, and $\Sigma_{\perp}(\bm k)$ are canceled by the renormalization of the chemical potential, Fermi velocity, and energy gap, respectively. The explicit expressions of the counterterms, which  are consistent with previous results from Refs.~\cite{gonzalez1994non,kotov2008polarization}
are given below for the two separate cases of e-e interaction.

We remark on a notable feature of the Hall response. Because of the topological nature of the Hall response, i.e. the fact that $\chi_{j_{x}j_{y}}^{(0)}$ is independent of $\vf$, no additional first-order counterterm diagram is produced from $\chi_{j_{x}j_{y}}^{(0)}$ as a consequence of the renormalization $\vf\to\vf+\delta\vf\ct$. At the same time, there is no divergence to first order that would have to be canceled by such a counterterm diagram. This is precisely due to the cancellation of the $\bm \Sigma_{\parallel}$ terms between self-energy and vertex diagrams. This in turn implies that the renormalization of the chemical potential and band gap are sufficient to ensure the finiteness of first-order response.

For the case of a contact potential $V_{c}$, we find that the counterterms are~[see Eqs.~(\ref{eq:sup_Sigma_0_finite_Vc})--(\ref{eq:sup_Sigma_perpendicular_finite_Vc})]
\begin{gather}
\frac{\delta\mu\ct}{\mu} =- \frac{N_{0}V_{c}}{4\,\mbox{sinc}(2\pi/n)}\frac{\Delta}{\mu}\left(\frac{\vf\Lambda}{\Delta}\right)^{2},\\
\frac{\delta\vf\ct}{\vf} = -\frac{N_{0}V_{c}}{4\,\mbox{sinc}(\pi/n)}\left(\frac{\vf\Lambda}{\Delta}\right),\\
\frac{\delta\Delta\ct}{\Delta} = -\frac{N_{0}V_{c}}{2}\left(\frac{\vf\Lambda}{\Delta \,\mbox{sinc}(\pi/n)}-1\right),
\end{gather}
where \(N_{0}=\Delta/(2\pi \vf^{2}) \) is the density of states (at zero temperature) at the bottom of the conduction band.

Thus, after some lengthy algebra~(see Appendix \ref{appendix_C}), we find the contribution from the first-order diagrams

\begin{align}
\label{eq:total_contribution_Vc}
\sigma_{H}^{(1)}=&\frac{\bar{\sigma}_{H}}{2}[\FD_{-2}^{+}(1-\FD_{-2}^{+}-2\FD_{0}^{+})-2 \FD_{1}^{-}\FD_{-3}^{-}\nn
&+\FD_{0}^{+}(\fh(\Delta)+\fe(\Delta)+1)
-\FD_{1}^{-}(\fh(\Delta)-\fe(\Delta))].\nn
\end{align}

This equation is one of the central results of our paper. We reintroduced the factors of $\hbar$ and defined $\bar{\sigma}_{H}= N_{0}V_{c}~ e^{2}/2h $ as our effective coupling constant multiplied by one fourth of the conductance quantum (since we neglect spin and valley degeneracies).

\begin{figure}[t!]
    \centering
    \includegraphics{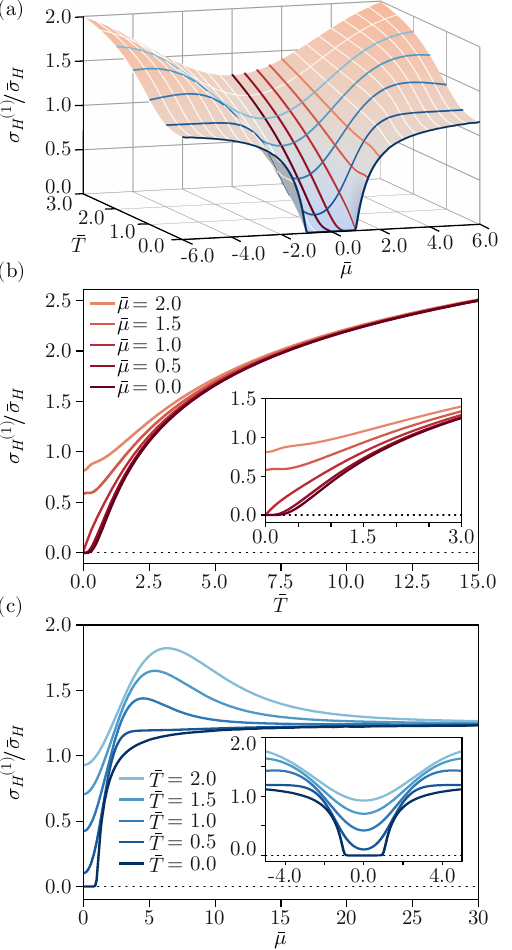}
    \caption{(a) The first-order correction due to contact e-e interactions to the Hall conductivity $\sigma_{H}^{(1)}$ as a function of both chemical potential and temperature. (b) The correction $\sigma_{H}^{(1)}$ as a function of temperature, for various chemical potentials, along the cuts shown in the same colours in (a). A logarithmic divergence at high temperatures can be observed. (c) The correction $\sigma_{H}^{(1)}$ as a function of the chemical potential, along the cuts shown in the same colours in (a), for fixed values of the temperature. At large doping, curves converge to the value of $5/4$. (Insets) Magnifications of the main plots in a similar range of values as the 3D plot.}
    \label{fig:2}
\end{figure}

We evaluate Eq. \eqref{eq:total_contribution_Vc} numerically as a function of both chemical potential and temperature. We start by discussing the case when the factor $N_{0}V_{c}$ is constant, i.e. it is independent of temperature and chemical potential. In this case, the behavior of the first-order correction to the Hall conductivity as a function of doping and temperature is shown in Fig. \ref{fig:2}. The particle-hole symmetry is clearly reflected in the fact that the correction is an even function of the chemical potential. 
An important feature of the Hall response that can be inferred from Fig.~\ref{fig:2}(a)  is the fact that the first-order correction vanishes at $\bar{T}=0$ when the chemical potential is placed inside the gap, i.e. $|\bar{\mu}| \leq 1$. This is not by chance, but it is a consequence of the Coleman--Hill theorem~\cite{coleman1985no}, which states that there are no perturbative corrections to the topological part of the Hall conductivity. 
It is worth noting that the validity of this theorem for our model is not a trivial statement, since the original proof assumed Lorentz invariance while nonretarded e-e interactions break Lorentz invariance~\footnote{This is because the Coulomb interaction is mediated by photons, which are much faster than the interacting electrons, thus making the interactions \enquote{instantaneous}.}. The theorem was extended to nonretarded interactions between electrons described by tight-binding models in Ref.~\cite{zhang2020influenceanomalous}.

Figure \ref{fig:2}(b) shows the temperature dependence of the correction $\sigma_{H}^{(1)}$ for a set of values of the chemical potential. All curves have the same asymptotic behavior, which highlights the logarithmic divergence at high temperatures, i.e.
\( \sigma_{H}^{(1)}\to\ln 2 (\ln \bar{T} +\mathcal{C})\bar{\sigma}_{H}\),
where $\mathcal{C}\approx 0.98$. 

Figure \ref{fig:2}(c) shows $\sigma_{H}^{(1)}$ as a function of chemical potential for fixed values of temperature. The curve at $\bar{T} = 0$ clearly proves the validity of the Coleman--Hill theorem, i.e. the vanishing of the correction to the Hall conductivity within the band gap. In the
Fermi liquid regime (i.e. for small temperatures and $|\bar{\mu}|>1$) the correction approaches the limit \(\sigma_H^{(1)} \to (5\bar{\mu}^{2}-2|\bar{\mu}|-3)/(4\bar{\mu}^{2})\bar{\sigma}_{H}\).

\begin{figure}[t!]
    \centering
    \includegraphics{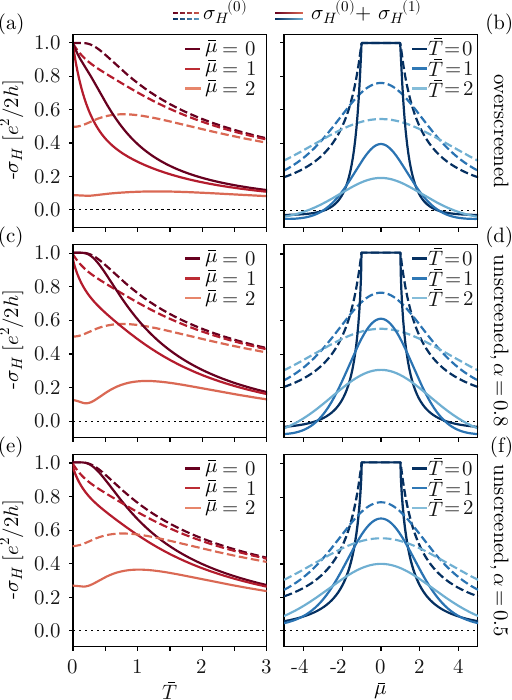}
    \caption{A comparison between the noninteracting ($\sigma_{H}^{(0)}$, dashed lines) and full ($\sigma_{H}=\sigma_{H}^{(0)}+\sigma_{H}^{(1)}$, solid lines) Hall conductivities. Note that an overall minus sign was introduced. (a) Curves are shown as a function of temperature for fixed values of the chemical potential for a heavily screened Coulomb interaction. (b) Curves are shown as a function of chemical potential for fixed values of temperature for a heavily screened Coulomb interaction. [(c),(d)] Same as (a) and (b), respectively, for an unscreened Coulomb interaction with an effective fine structure constant $\alpha = 0.8$. [(e),(f)] Same as (a) and (b), respectively, for an unscreened Coulomb interaction with $\alpha = 0.5$.}
    \label{fig:3}
\end{figure}
\subsection{Overscreened Coulomb interactions}
When the density of states is large, the interaction is heavily screened. For a static Thomas--Fermi screening~\cite{sarma2011electronic}, the contact potential becomes $V_c \approx \nu^{-1}$, where $\nu (\mu,T)= \partial_{\mu} n(\mu,T)$ is the thermodynamic density of states. The effective coupling constant $N_{0}V_{c}$ becomes
\be
\label{eq:N0Vc}
N_{0}V_{c}=4\bar{T}\left[ \int_{|x|>1} dx |x| \cosh^{-2}\left(\frac{x-\bar{\mu}}{2\bar{T}}\right)\right]^{-1}.
\ee

The Hall conductivity to first order in interaction, $\sigma_{H}=\sigma_{H}^{(0)}+\sigma_{H}^{(1)}$, is shown with solid lines in Figs.~\ref{fig:3}(a) and \ref{fig:3}(b). Comparing it with the noninteracting result (dashed lines), we see that e-e interactions in the high screening regime can be strong enough to offset the noninteracting Hall response: $\sigma_{H}$ vanishes quicker than $\sigma_{H}^{(0)}$ with increasing chemical potential or temperature.
This effect is similar to that of white-noise disorder, which strongly suppresses the anomalous Hall conductivity at high electron density~\cite{ado2017sensitivity}.

Figure \ref{fig:3}(b) also shows that the conductivity can become negative at large enough chemical potentials. This feature is most likely an artifact of the truncation to first-order in perturbation theory, since the choice $V_{c}\approx \nu^{-1}$ does not allow the potential to be arbitrarily small. We thus expect that summing higher order terms would remedy this feature.
Note that even though Eq.~\eqref{eq:N0Vc} is formally valid only when the carrier density is large, we still continued the dark-blue-solid line ($\bar{T}=0$) into the interval $|\bar{\mu}|\leq1$ in Fig. \ref{fig:3}(b) for completeness. This can be done because, due to the Coleman--Hill theorem, $\sigma_{H}^{(1)}$ is always zero in that region, irrespective of the exact form of $V_{c}$. A similar reasoning holds for the dark-red-solid line ($\bar{\mu}=0$) in Fig. \ref{fig:3}(a) at very low temperatures.

\subsection{Unscreened Coulomb interaction}

As before, UV divergences are isolated and absorbed into their corresponding counterterms~[Eqs.~(\ref{eq:sup_Sigma_0_Coulomb})--(\ref{eq:sup_Sigma_perpendicular_Coulomb})]
\begin{gather}
\frac{\delta\mu\ct}{\mu} =- \frac{\alpha}{2}\frac{\Delta}{\mu}\left(\frac{\vf\Lambda}{\Delta\,\mbox{sinc}(\pi/n)}\right),\\
\frac{\delta\vf\ct}{\vf} =- \frac{\alpha}{4}\left[\ln\left(\frac{\vf\Lambda}{\Delta}\right)+\ln2+1\right],\\
\frac{\delta\Delta\ct}{\Delta} =- \frac{\alpha}{2}\left[\ln\left(\frac{\vf\Lambda}{\Delta}\right)+\ln2\right],
\end{gather}
with $\alpha = e^2/(\kappa\hbar \vf)$ the effective fine structure constant.

The Hall conductivity as a function of temperature and chemical potential is shown in Figs.~\ref{fig:3}(c)--\ref{fig:3}(f). The exact expression for $\sigma_{H}$ can be found in Appendix~\ref{appendix_D}.

In order to compare these results with those obtained for the high screening regime, we have chosen the value of the effective fine structure constant to be $\alpha=0.8$ [panels (c)--(d)] and $\alpha=0.5$ [panels (e)--(f)]. Similar to Figs.~\ref{fig:3}(a)--\ref{fig:3}(b), in both cases $\sigma_{H}$ (solid lines) decays faster than $\sigma_{H}^{(0)}$ (dashed lines) as the temperature and the chemical potential increase. In Figs.~\ref{fig:3}(c) and \ref{fig:3}(d) we can see that choosing $\alpha=0.8$ gives a result of striking similarity with the high-screening regime. For $|\bar{\mu}|\gtrsim 3$, the Hall conductivity becomes negative, thus exhibiting the same artifact as Fig.~\ref{fig:3}(b).

Picking a smaller value of the coupling constant, e.g. $\alpha=0.5$, moves the spurious zero of $\sigma_{H}$ to larger values of the chemical potential, as can be seen in Fig.~\ref{fig:3}(f). This indicates that Hall conductivity becoming negative is most likely an artifact of truncation to first order.

\section{Discussion and conclusions}

We calculated the first-order correction to the Hall conductivity due to e-e interactions. Firstly, we showed that the topological part of the Hall conductivity is robust against interactions, in agreement with the Coleman--Hill theorem. Secondly, we found that the effect of interactions can be strong enough to offset the noninteracting contribution. As a result, the total Hall conductivity goes to zero faster than the noninteracting one when either the temperature or the chemical potential increase. These features are present for both heavily screened and unscreened e-e interactions.

Note that this calculation pertains to the continuum model of a single Dirac cone. In a lattice model, massive Dirac fermions can be found as low-energy quasiparticle excitations of, e.g., a two-dimensional hexagonal lattice with broken sub-lattice symmetry. In this case, massive Dirac fermions of masses $\Delta_{K}$ and $\Delta_{K^{\prime}}$ and opposite chiralities would appear in two distinct points of the Brillouin zone, $K$ and $K^\prime$ respectively. Then the total Hall conductivity is the sum of the Hall conductivities of the two fermion species. This implies that, if the system is invariant under time-reversal symmetry, such that $\Delta_{K}=\Delta_{K^{\prime}}$, then $\sigma_{H}=\sigma_{H}^{(0)}+\sigma_{H}^{(1)}$ exactly cancels between the two valleys. However, if time-reversal symmetry is broken but spatial inversion symmetry is preserved, such that $\Delta_{K}=-\Delta_{K^{\prime}}$, then they add up to a finite value.

It is important to point out that the two valleys can be treated separately. This is because intervalley Coulomb backscattering is suppressed by a large intervalley momentum transfer $|{\bm K}-{\bm K}^{\prime}|$ and therefore can be safely neglected. This remains true also for the contact interaction which should be interpreted as a heavily screened Coulomb potential. In fact, the interaction appears as contact only at length scales larger than the screening length. On the other hand, on smaller scales it recovers the original unscreened Coulomb interaction. We stress that this is consistent with our choice of regularization scheme, which forces the potential to decay when the momentum transfer exceeds the cut-off value $\Lambda$, which is taken to be much smaller than $|{\bm K} - {\bm K}'|$. Thus, intervalley backscattering can always be neglected.

Possible experimental platforms where these results can be tested include any gapped Dirac material with broken time-reversal symmetry. For instance, a gap can be opened in graphene via substrate effects~\cite{zhou2007substrate}, while creating an imbalance in the population of the opposite valleys by exposing it to circularly polarized light~\cite{mciver2020light}. Moreover, a gap can also be induced by placing graphene in a chiral cavity~\cite{wang2019cavity} or by proximity coupling it to a ferromagnet~\cite{wang2015proximity,hogl2020quantum,zollner2016theory}. We note that if the gap originates from the coupling to a ferromagnet, then the magnitude of the gap can decrease with temperature together with the magnetization in the substrate, which may make the Hall conductivity decay even faster (depending on the value of the Curie temperature) than simply due to e-e interactions. As an alternative to graphene, the effect could also be observed in certain transition-metal dichalcogenides~\cite{habe2017anomalous, cao2012valley}, which have the advantage of a pre-existing gap due to spin-orbit coupling, and at the surfaces of 3D topological insulators~\cite{chen2010massive,bhattacharyya2021recent}.

We stress that our results are applicable to other effects beyond anomalous Hall transport. We expect similar corrections to exist in, e.g., spin-Hall~\cite{sinova2015spin,kane2005quantum} or orbital-Hall~\cite{choi2023observation,bhowal2021orbital} conductivities of sufficiently clean materials, although more work is needed to clarify the impact of e-e interactions on these effects.

\begin{acknowledgments}
A.D.D.-I. and D.-A.D. acknowledge support from the Engineering and Physical Sciences Research Council, Grant No. EP/T517823/1. A.P. and A.E.K. acknowledge support from the Leverhulme Trust under the Grant Agreement No. RPG-2019-363. The authors also acknowledge support from the European Commission under the EU Horizon 2020 MSCA-RISE-2019 programme (Project No. 873028 HYDROTRONICS).
\end{acknowledgments}

\allowdisplaybreaks
\appendix
\section{\label{appendix_A}Zeroth-order diagram}

In the reverse order of the arrows in the diagram from Fig.~\ref{fig:1}(a), we use the finite-temperature Feynman rules given in  Sec. 6.4 in Ref.~\cite{vignale2005quantum}\footnote{Note that there is a wrong minus sign in Rule (vii), which was later corrected in the errata of this book.} to calculate the noninteracting current-current response function
\begin{align}
\chi_{j_{x}j_{y}}^{(0)}(\bm q, i\omega)=&\frac{1}{\beta}\sum_{i\omega_{m}}\int_{\bm k} \tr [G^{(0)}( \bm k, i\omega_{m})j_{x}\nn
&\times G^{(0)}( \bm k+\bm q, i\omega_{m}+i\omega)j_{y}]\nn
&=-\omega\frac{\vf^{2}\Delta}{2}\int_{\bm k} \frac{\fe(-\vare_{k})-\fe(\vare_{k})}{\vare_{k}^{3}}.
\end{align}
In this equation we used the shorthand notation $\int_{\bm k}=\int d\bm k/(2\pi)^{2}$ for the integral, and $\sum_{i\omega_{m}}$ for the sum over Matsubara frequencies. The key advantage for evaluating this sum lies in the fact that the Fermi--Dirac distribution function $\fe(z) =[e^{\beta z}+1]^{-1}$ has its poles at fermionic Matsubara frequencies $i\omega_{m}=i(2m + 1)\pi/\beta$, where $m$ is an integer, and the corresponding residues are $\mathrm{Res}[\fe(z)]=-1/\beta$. For a function of known simple poles, $g(z)=\Pi_{j}(z-z_{j})^{-1}$,
\be
\label{eq:sup_Matsubara_sum_simple_poles}
\frac{1}{\beta}\sum_{i\omega_{m}}g(i\omega_{m})=\sum_{j}\mathop{\mathrm{Res}}_{\,\,z = z_j}
\big[g(z)\big]\fe(z_{j}).
\ee

After performing the Wick rotation to real frequencies $i\omega\to\omega+i0^+$ and using the Kubo formula given in Eq.~\eqref{eq:kubo}, we obtain the Hall conductivity
\be
\label{eq:sup_sigma_H0}
\sigma_{H}^{(0)}(\mu, T)&=&-\frac{e^{2}\Delta}{4\pi}\int_{\Delta}^{\infty}d\vare_{k}\frac{\fe(-\vare_{k})-\fe(\vare_{k})}{\vare_{k}^{2}}~.
\ee

Using the Fermi--Dirac-like integral defined before in Eq.~\eqref{eq:fermi_dirac_integral} and reintroducing the factors of $\hbar$, we finally get
\be
\sigma_{H}^{(0)}(\bar{\mu}, \bar{T})&=&-\frac{e^{2}}{2h}\left(1-\FD_{-2}^{+}(\bar{\mu},\bar{T})\right)~.
\ee

At zero temperature, we can take the limit of $\FD_{n}^{\pm}$,
\begin{gather}
\label{eq:sup_Fn+}
\FD_{n}^{+}(\bar{\mu},0)=\int_{1}^{|\bar{\mu}|}dx \,x^{n}=~\frac{|\bar{\mu}|^{n+1}-1}{n+1}~,\\
\label{eq:sup_Fn-}
\FD_{n}^{-}(\bar{\mu},0)=\text{sign}(\bar{\mu})\int_{1}^{|\bar{\mu}|}dx \,x^{n}=\text{sign}(\bar{\mu})\frac{|\bar{\mu}|^{n+1}-1}{n+1}~,\nn
\end{gather}
which are true for $|\bar{\mu}|>1$, but $\FD_{n}^{\pm}(|\bar{\mu}|\leq1,0)=0$ when the chemical potential lies inside the gap.

It can easily be seen that in this limit we get $\sigma_{H}^{(0)}=-e^{2}/2h$ when $|\bar{\mu}|\leq1$ and $\sigma_{H}^{(0)}(\bar{\mu},0)=-(e^{2}/2h)(1/|\bar{\mu}|)$ otherwise. This recovers the well-known result for the anomalous Hall conductivity from Ref.~\cite{sinitsyn2006charge}.

\widetext

\section{\label{appendix_B}First-order diagrams}

Using the same Feynman rules for the other three diagrams, i.e. Figs.~\ref{fig:1}(b)--\ref{fig:1}(d), we calculate the response functions
\begin{gather}
\chi_{j_{x}j_{y}}^{\text{EX}}(0, i\omega)=-\frac{1}{\beta^{2}}\!\!\sum_{i\omega_{n},i\omega_{n}\pr}\!\!\int_{\bm k}\int_{\bm k\pr} \!\!V(\bm k-\bm k\pr)\tr [j_{x}G^{(0)}( \bm k, i\omega_{n}+i\omega)G^{(0)}( \bm k\pr, i\omega_{n}\pr+i\omega)j_{y}G^{(0)}( \bm k\pr, i\omega_{n}\pr)G^{(0)}( \bm k, i\omega_{n})],\\
\chi_{j_{x}j_{y}}^{\text{SE}1}(0, i\omega)=\frac{1}{\beta}\sum_{i\omega_{n}}\int_{\bm k} \tr [j_{x}G^{(0)}( \bm k, i\omega_{n}+i\omega)\Sigma(\bm k )G^{(0)}( \bm k, i\omega_{n}+i\omega)j_{y}G^{(0)}( \bm k, i\omega_{n})]\!,\\
\chi_{j_{x}j_{y}}^{\text{SE}2}(0, i\omega)=\frac{1}{\beta}\sum_{i\omega_{n}}\int_{\bm k} \tr [j_{x}G^{(0)}( \bm k, i\omega_{n}+i\omega)j_{y}G^{(0)}( \bm k, i\omega_{n})\Sigma(\bm k)G^{(0)}( \bm k, i\omega_{n})]\!.
\end{gather}

The first one is the easiest to evaluate, and it can be simplified by using a parity transformation $\bm k \to - \bm k$ and $\bm k\pr \to - \bm k\pr$. The latter two, however, have second-order poles, which need to be treated carefully. These poles give rise to derivatives of the occupation functions, and change Eq.~\eqref{eq:sup_Matsubara_sum_simple_poles} into
\be
\label{eq:sup_matsubara_sums_second_order}
\frac{1}{\beta}\sum_{i\omega_{m}}g(i\omega_{m})=\!\!\!\!\sum_{j\in\{j_{1},j_{2}\}}\!\!\!\mathop{\mathrm{Res}}_{\,\,z = z_j}
\big[g(z)\big]\fe(z_{j})+\sum_{j_{2}} \Bigg[\frac{d\fe(z)}{dz}\Big((z-z_{j_{2}})^{2}g(z)\Big)\Bigg]_{z=z_{j_{2}}}~,
\ee
where $j_{1}$ and $j_{2}$ are the simple and double poles, respectively.

After summing over the Matsubara frequencies and performing the traces over the Pauli matrices, we find that the first-order current-current response functions are
\be
\label{eq:sup_exchange_diagram}
\chi_{j_{x}j_{y}}^{\text{EX}}(0,i\omega)=-\omega \frac{\vf^{2}\Delta}{4}\int_{\bm k}\int_{\bm k\pr}
V(\bm k-\bm k\pr)\Bigg[\frac{\vare_{k\pr}^{2}+\Delta^{2}}{\vare_{k\pr}^3\vare_{k}^3}+\frac{ \vf^{2}\bm k\pr\cdot\bm k}{\vare_{k\pr}^3\vare_{k}^3}\Bigg]\Phi^{-}_{k\pr}\Phi^{-}_{k},
\ee
\begin{align}
\label{eq:sup_self_energy_diagram}
\chi_{j_{x}j_{y}}^{\text{SE}}&=\omega \frac{\vf\Delta}{2}\!\int_{\bm k}\!\Bigg\{\Sigma_{0}(\bm k)\Bigg[\frac{\vf}{\vare_{k}^{3}}\frac{\partial \Phi^{+}_{k}}{\partial \vare_{k}}\Bigg]\!+\!\bm \Sigma_{\parallel}(\bm k)\cdot\frac{\vf^{2}\bm k}{\vare_{k}}\Bigg[\frac{3}{\vare_{k}^{4}}\Phi^{-}_{k}-\frac{1}{\vare_{k}^{3}} \frac{\partial \Phi^{-}_{k}}{\partial \vare_{k}}\Bigg]\!+\!\Sigma_{\perp}(\bm k)\frac{\vf}{\Delta}\Bigg[\frac{3\Delta^{2}-\vare_{k}^{2}}{\vare_{k}^{5}}\Phi^{-}_{k}-\frac{\Delta^{2}}{\vare_{k}^{4}}\frac{\partial \Phi^{-}_{k}}{\partial \vare_{k}}\Bigg]\!\Bigg\}\nn
&=\omega \frac{\vf\Delta}{2}\!\int_{\bm k}\!\Bigg\{\Sigma_{0}(\bm k)\Bigg[\frac{\vf}{\vare_{k}^{3}}\frac{\partial \Phi^{+}_{k}}{\partial \vare_{k}}\Bigg]\!-\!
\bm \Sigma_{\parallel}(\bm k)\cdot\frac{\vf^{2}\bm k}{\vare_{k}}\frac{\partial}{\partial \vare_{k}}\!\left(\frac{\Phi^{-}_{k}}{\vare_{k}^{3}}\!\right)\!+\!\Sigma_{\perp}(\bm k)\frac{\vf}{\Delta}\Bigg[\!\left(\frac{2}{\vare_{k}^{3}}-\frac{3}{\vare_{k}^{4}}\frac{\vf^{2}k^{2}}{\vare_{k}}\right)\!\Phi^{-}_{k}\!+\!\left(\frac{\vf^{2}k^{2}}{\vare_{k}^{4}}-\frac{1}{\vare_{k}^{2}}\right)\!\frac{\partial \Phi^{-}_{k}}{\partial \vare_{k}}\Bigg]\!\Bigg\}\nn
&= \omega \frac{\vf\Delta}{2}\!\int_{\bm k}\!\Bigg\{\Sigma_0(\bm k)\Bigg[\frac{\vf}{\vare_{k}^{3}}\frac{\partial \Phi^{+}_{k}}{\partial\vare_{k}}\Bigg]\!-\!\bm \Sigma_{\parallel}(\bm k)\cdot \Bigg[\bm \nabla_{\bm k}^{\text{2D}}\left(\frac{\Phi^{-}_{k}}{\vare_{k}^{3}}\right)\Bigg]\!+\!\Sigma_{\perp}(\bm k)\Bigg[\frac{\vf}{\Delta}\bm \nabla_{\bm k}^{\text{2D}}\cdot \left( \bm k \frac{\Phi^{-}_{k}}{\vare_{k}^{3}}\right)-\frac{\vf}{\vare_{k}^{2}\Delta}\frac{\partial \Phi^{-}_{k}}{\partial \vare_{k}}\Bigg]\!\Bigg\},
\end{align}
where $\Sigma_{0}$ is given in Eq.~\eqref{eq:sigma_0} and we split $\bm \Sigma$ from Eq.~\eqref{eq:sigma_bold} into $\bm \Sigma_{\parallel}=(\Sigma_{x},\Sigma_{y})$ and $\Sigma_{\perp}= \Sigma_{z}$. When integrating the above by parts, $\bm \nabla_{\bm k}^{\text{2D}}$ gives both vanishing boundary terms and the 2D divergence (gradient) of $\bm \Sigma_{\parallel}(\Sigma_{\perp})$,
\be
\bm \nabla_{\bm k}^{\text{2D}} \cdot \bm \Sigma_{\parallel}(\bm k) &=& \bm \nabla_{\bm k}^{\text{2D}} \cdot \left(\frac{1}{2} \int_{\bm k\pr} \frac{\bm d_{\parallel}(\bm k\pr)}{\vare_{k\pr}} V(\bm k - \bm k\pr) \Phi_{k\pr}^{-}\right)\nn
&=& \frac{1}{2} \int_{\bm k\pr} \left[\bm \nabla_{\bm k\pr}^{\text{2D}} \cdot \left(\frac{\bm d_{\parallel}(\bm k\pr)}{\varepsilon_{k\pr}} \right) \Phi_{k\pr}^{-}+ \frac{\bm d_{\parallel}(\bm k\pr)}{\varepsilon_{k\pr}} \cdot \bm \nabla_{\bm k\pr}^{\text{2D}} \left(\Phi_{k\pr}^{-} \right) \right] V(\bm k - \bm k\pr)\nn
&=& \frac{\vf}{2} \int_{\bm k\pr} \left[\left(\frac{\varepsilon_{k\pr}^{2}+\Delta^{2}}{\varepsilon_{k\pr}^{3}} \right) \Phi_{k\pr}^{-}+ \frac{\bm k\pr}{\varepsilon_{k\pr}} \cdot \bm \nabla_{\bm k\pr}^{\text{2D}} \left(\Phi_{k\pr}^{-} \right) \right] V(\bm k - \bm k\pr),
\ee
\be
\bm \nabla_{\bm k}^{\text{2D}}  \Sigma_{\perp}(\bm k) &=& \bm \nabla_{\bm k}^{\text{2D}} \left( \frac{1}{2} \int_{\bm k\pr} \frac{\Delta}{\vare_{k\pr}} V(\bm k - \bm k\pr) \Phi_{k\pr}^{-}\right)\nn
&=& \frac{\Delta}{2} \int_{\bm k\pr} \left[\bm \nabla_{\bm k\pr}^{\text{2D}} \left(\frac{1}{\varepsilon_{k\pr}} \right) \Phi_{k\pr}^{-}+ \frac{1}{\varepsilon_{k\pr}} \bm \nabla_{\bm k\pr}^{\text{2D}} \left(\Phi_{k\pr}^{-} \right) \right] V(\bm k - \bm k\pr)\nn
&=& \frac{\Delta}{2} \int_{\bm k\pr} \left[\left(-\frac{\vf^{2}\bm k\pr}{\varepsilon_{k\pr}^{3}} \right) \Phi_{k\pr}^{-}+ \frac{1}{\varepsilon_{k\pr}}\bm \nabla_{\bm k\pr}^{\text{2D}} \left(\Phi_{k\pr}^{-} \right) \right] V(\bm k - \bm k\pr).
\ee

Plugging these expressions back into Eq.~\eqref{eq:sup_self_energy_diagram} gives an exact cancellation with the exchange response function
\begin{align}
\chi_{j_{x}j_{y}}^{\text{SE}}(0,i\omega)=&-\omega \frac{\vf\Delta}{2}\int_{\bm k}\Bigg\{-\Sigma_0(\bm k)\Bigg[\frac{\vf}{\vare_{k}^{3}}\frac{\partial \Phi^{+}_{k}}{\partial\vare_{k}}\Bigg]-\bm \nabla_{\bm k}^{\text{2D}} \cdot \bm \Sigma_{\parallel}(\bm k)\left(\frac{\Phi^{-}_{k}}{\vare_{k}^{3}}\right)+ \bm \nabla_{\bm k}^{\text{2D}} \Sigma_{\perp}(\bm k)\cdot \left( \frac{\vf \bm k}{\Delta} \frac{\Phi^{-}_{k}}{\vare_{k}^{3}}\right)+\Sigma_{\perp}(\bm k) \frac{\vf}{\vare_{k}^{2}\Delta}\frac{\partial \Phi^{-}_{k}}{\partial \vare_{k}}\Bigg\}\nn
=&-\omega \frac{\vf\Delta}{2}\int_{\bm k}\Bigg\{-\Sigma_0(\bm k)\Bigg[\frac{\vf}{\vare_{k}^{3}}\frac{\partial \Phi^{+}_{k}}{\partial\vare_{k}}\Bigg]+\Bigg[\Sigma_{\perp}(\bm k) \frac{\vf}{\vare_{k}^{2}\Delta}\frac{\partial \Phi^{-}_{k}}{\partial \vare_{k}}\Bigg]-\frac{1}{2}\int_{\bm k\pr}V(\bm k - \bm k\pr)\frac{\vf \bm k\pr}{\vare_{k}^{3}\vare_{k\pr}}\cdot \bm \nabla_{\bm k\pr}^{\text{2D}}(\Phi_{k\pr}^{-})\Phi_{k}^{-}\nn
&+\frac{1}{2}\int_{\bm k\pr}V(\bm k -\bm k\pr)\frac{\vf \bm k}{\vare_{k}^{3}\vare_{k\pr}}\cdot \bm \nabla_{\bm k\pr}^{\text{2D}}(\Phi_{k\pr}^{-})\Phi_{k}^{-}\Bigg\}+\underbrace{\omega \frac{\vf^{2}\Delta}{4}\int_{\bm k}\int_{\bm k\pr}V(\bm k-\bm k\pr)\Bigg[\frac{\vare_{k\pr}^{2}+\Delta^{2}+\vf^{2}\bm k \cdot \bm k\pr}{\vare_{k}^{3}\vare_{k\pr}^{3}}\Bigg]\Phi_{k}^{-}\Phi_{k\pr}^{-}}_{=-\chi_{j_{x}j_{y}}^{\text{EX}}(0,i\omega)},
\end{align}
which is in accordance with the Ward identity. Summing the two leaves us with the result previously given in Eq.~\eqref{eq:all_diagrams}.

We isolate the divergences in $\Sigma_0$, $\bm \Sigma_{\parallel}$, and $\Sigma_{\perp}$ by separating the contribution of the holes (in the lower band) from that of the Fermi sea, \textit{i.e.}, using the fact that \(\Phi^{\pm}_{k}
=\fe(-\vare_{k})\pm \fe(\vare_{k})=1-(\fh(\vare_{k})\mp \fe(\vare_{k}))\). Since both $\fe(\vare_k)$ and $\fh(\vare_k)$ are exponentially suppressed at large momentum, only the terms containing the unity are UV divergent 
\be
\label{eq:sup_Sigma_0}
\Sigma_0(\bm k)&=&-\frac{1}{2}\int_{\bm k\pr} V(\bm k-\bm k\pr) \Phi_{k\pr}^{+}=-\frac{1}{2}\int_{\bm k\pr} V(\bm k-\bm k\pr) (1-\fh(\vare_{k\pr})+\fe(\vare_{k\pr}))\nn
&=&-\frac{1}{2}\int_{\bm k\pr} V(\bm k - \bm k\pr)-\frac{1}{2}\int_{\bm k\pr} V(\bm k-\bm k\pr) (\fe(\vare_{k\pr})-\fh(\vare_{k\pr})),
\ee
\be
\label{eq:sup_Sigma_parallel}
\bm \Sigma_{\parallel}(\bm k)&=&\frac{1}{2}\int_{\bm k\pr} V(\bm k-\bm k\pr) \frac{\bm d_{\parallel}(\bm k\pr)}{\vare_{k\pr}}\Phi_{k\pr}^{-}=\frac{1}{2}\int_{\bm k\pr} V(\bm k-\bm k\pr) \frac{\vf\bm k\pr}{\vare_{k\pr}}(1-\fh(\vare_{k\pr})-\fe(\vare_{k\pr}))\nn
&=&\frac{\vf}{2}\int_{\bm k\pr} V(\bm k - \bm k\pr) \frac{\bm k\pr}{\vare_{k\pr}}-\frac{1}{2}\int_{\bm k\pr} V(\bm k-\bm k\pr) \frac{\vf \bm k\pr}{\vare_{k\pr}}(\fe(\vare_{k\pr})+\fh(\vare_{k\pr})),
\ee
\be
\label{eq:sup_Sigma_perpendicular}
\Sigma_{\perp}(\bm k)&=&\frac{1}{2}\int_{\bm k\pr} V(\bm k-\bm k\pr) \frac{\Delta}{\vare_{k\pr}}\Phi_{k\pr}^{-}=\frac{1}{2}\int_{\bm k\pr} V(\bm k-\bm k\pr) \frac{\Delta}{\vare_{k\pr}}(1-\fh(\vare_{k\pr})-\fe(\vare_{k\pr}))\nn
&=&\frac{\Delta}{2}\int_{\bm k\pr} V(\bm k - \bm k\pr) \frac{1}{\vare_{k\pr}}-\frac{1}{2}\int_{\bm k\pr} V(\bm k - \bm k\pr)\frac{\Delta}{\vare_{k\pr}}(\fe(\vare_{k\pr})+\fh(\vare_{k\pr})).
\ee

The divergences introduced by $\Sigma_0$, $\bm \Sigma_{\parallel}$, and $\Sigma_{\perp}$ become finite after the renormalization of the chemical potential, Fermi velocity and energy gap, respectively. Since these divergences depend on the form of the interaction, we treat the case of a constant contact potential $V(\bm k)=V_{c}$ and that of a Coulomb potential $V(\bm k)=2\pi e^2/(\kappa k)$ separately.

\section{\label{appendix_C}Constant contact potential}

We consider a regularised potential of the form $V(\bm k)\to V_{c}\left[1+(|\bm k|/\Lambda)^{n}\right]^{-1}$, where $n$ is an integer. Introducing the counterterms in the previous equations leaves us with 
\be
\label{eq:sup_Sigma_0_finite_Vc}
\Sigma_0(\bm k)-\delta\mu\ct
&=&-\frac{1}{2}\int_{\bm k\pr} V(\bm k\pr)-\frac{1}{2}\int_{\bm k\pr} V(\bm k-\bm k\pr) (\fe(\vare_{k\pr})-\fh(\vare_{k\pr}))-\delta\mu\ct\nn
&=&-\frac{V_{c}\Lambda^{2}}{8\pi\,\mbox{sinc}(2\pi/n)}-\delta\mu\ct-\frac{1}{2}\int_{\bm k\pr} V(\bm k-\bm k\pr) (\fe(\vare_{k\pr})-\fh(\vare_{k\pr}))\nn
&=&-\mu\underbrace{\left[\frac{N_{0}V_{c}}{4\,\mbox{sinc}(2\pi/n)}\frac{\Delta}{\mu}\left(\frac{\vf\Lambda}{\Delta}\right)^{2}+\frac{\delta\mu\ct}{\mu}\right]}_{=0}-\frac{1}{2}\int_{\bm k\pr} V(\bm k-\bm k\pr) (\fe(\vare_{k\pr})-\fh(\vare_{k\pr})),
\ee
\be
\label{eq:sup_Sigma_parallel_finite_Vc}
\bm \Sigma_{\parallel}(\bm k)+\delta\vf\ct\bm k
&=&\frac{1}{2}\int_{\bm k\pr} V(\bm k\pr) \frac{\vf(\bm k+ \bm k\pr)}{\vare_{\bm k + \bm k\pr}}-\frac{1}{2}\int_{\bm k\pr} V(\bm k-\bm k\pr) \frac{\vf \bm k\pr}{\vare_{k\pr}}(\fe(\vare_{k\pr})+\fh(\vare_{k\pr}))+\delta\vf\ct\bm k\nn
&=&\frac{\vf}{2}\int_{\bm k\pr} V(\bm k\pr) \left(\cancel{\frac{\bm k\pr}{\vare_{k\pr}}}+\bm k \frac{\vare_{k\pr}^{2}+\Delta^{2}}{2\vare_{k\pr}^{3}}+\mathcal{O}(k^{2})\right)-\frac{1}{2}\int_{\bm k\pr} V(\bm k-\bm k\pr) \frac{\vf \bm k\pr}{\vare_{k\pr}}(\fe(\vare_{k\pr})+\fh(\vare_{k\pr}))+\delta\vf\ct\bm k\nn
&=&\bm k \left(\frac{V_{c}\Lambda}{8\pi\,\mbox{sinc}(\pi/n)}+\delta\vf\ct\right)-\frac{1}{2}\int_{\bm k\pr} V(\bm k-\bm k\pr) \frac{\vf \bm k\pr}{\vare_{k\pr}}(\fe(\vare_{k\pr})+\fh(\vare_{k\pr}))\nn
&=&\vf\bm k \underbrace{\left[\frac{N_{0}V_{c}}{4\,\mbox{sinc}(\pi/n)}\frac{\vf\Lambda}{\Delta}+\frac{\delta\vf\ct}{\vf}\right]}_{=0}-\frac{1}{2}\int_{\bm k\pr} V(\bm k-\bm k\pr) \frac{\vf \bm k\pr}{\vare_{k\pr}}(\fe(\vare_{k\pr})+\fh(\vare_{k\pr})),
\ee
\be
\label{eq:sup_Sigma_perpendicular_finite_Vc}
\Sigma_{\perp}(\bm k)+\delta\Delta\ct
&=&\frac{\Delta}{2}\int_{\bm k\pr} V(\bm k\pr) \frac{1}{\vare_{\bm k +\bm k\pr}}-\frac{1}{2}\int_{\bm k\pr} V(\bm k - \bm k\pr)\frac{\Delta}{\vare_{k\pr}}(\fe(\vare_{k\pr})+\fh(\vare_{k\pr}))+\delta\Delta\ct\nn
&=&\frac{\Delta}{2}\int_{\bm k\pr} V(\bm k\pr) \left(\frac{1}{\vare_{k\pr}}-\bm k \cdot \hspace{1.2mm}\cancel{\hspace{-1.2mm}\frac{\vf^2\bm k\pr}{\vare_{k\pr}^{3}}\hspace{-1.2mm}}\hspace{1.2mm}+\mathcal{O}(k^{2})\right)-\frac{1}{2}\int_{\bm k\pr} V(\bm k - \bm k\pr)\frac{\Delta}{\vare_{k\pr}}(\fe(\vare_{k\pr})+\fh(\vare_{k\pr}))+\delta\Delta\ct\nn
&=&\frac{V_c \Lambda}{4\pi\,\mbox{sinc}(\pi/n)}\frac{\Delta}{\vf}-\frac{V_{c}}{4\pi}\frac{\Delta^{2}}{\vf^{2}}+\delta\Delta\ct-\frac{\Delta}{2}\int_{\bm k\pr} V(\bm k-\bm k\pr) \frac{\fe(\vare_{k\pr})+\fh(\vare_{k\pr})}{\vare_{k\pr}}\nn
&=&\Delta\underbrace{\left[\frac{N_{0}V_c}{2\,\mbox{sinc}(\pi/n)} \frac{\vf\Lambda}{\Delta}-\frac{N_{0}V_{c}}{2}+\frac{\delta\Delta\ct}{\Delta}\right]}_{=0}-\frac{\Delta}{2}\int_{\bm k\pr} V(\bm k-\bm k\pr) \frac{\fe(\vare_{k\pr})+\fh(\vare_{k\pr})}{\vare_{k\pr}},
\ee
where the crossed terms vanish by rotational symmetry and integrals of all $\mathcal{O}(k^2)$ terms vanish identically in the limit $\Lambda\to\infty$.

Note that in addition to the divergence itself, we also absorbed a finite constant term into $\delta\Delta\ct$ in the last line. This is required to ensure that, after the renormalization, $\Delta$ is the \enquote{physical mass}, which is defined as the pole of the dressed Green’s function at zero momentum (at zero electron density). When $\mu=0$ and $T=0$, all the Fermi--Dirac distribution functions above vanish. Thus this requirement translates to $\Sigma_{\perp}(\bm k=0) + \delta\Delta\ct =0$. Choosing not to absorb this constant in the counterterm would make the renormalized mass related to the physical mass as $\Delta=\Delta^{\text{ph}}(1+N_{0}V_{c}/2)^{-1}$.

Plugging the (now finite) expressions for $\Sigma_{0}$ and $\Sigma_{\perp}$ from Eqs.~\eqref{eq:sup_Sigma_0_finite_Vc} and \eqref{eq:sup_Sigma_perpendicular_finite_Vc} into Eq.~\eqref{eq:all_diagrams}, we can perform the substitution $i\omega\to\omega+i0^+$, and use the Kubo formula to recover the Hall conductivity given in Eq.~\eqref{eq:total_contribution_Vc}.

When the temperature is zero and the chemical potential lies inside the gap $|\bar{\mu}|\leq1$, all functions $\FD^{\pm}_{n}$ are identically zero. However, when $|\bar{\mu}|>1$, we use the zero-temperature limits of $\FD_{n}^{\pm}$ given in Eqs.~\eqref{eq:sup_Fn+} and \eqref{eq:sup_Fn+}\eqref{eq:sup_Fn-} and the fact that $\fh(\Delta)+\fe(\Delta)\xrightarrow[]{T\to0}1$ and $\fh(\Delta)-\fe(\Delta)\xrightarrow[]{T\to0}-\text{sign}(\bar{\mu})$, to find the limit
\be
\sigma_{H}^{(1)}\xrightarrow[]{T\to0}\frac{5|\bar{\mu}|^{2}-2|\bar{\mu}|-3}{4|\bar{\mu}|^{2}}\bar{\sigma}_{H},
\ee
which at large doping converges to a finite value, i.e., \( \sigma_{H}^{(1)}\to5/4\, \bar{\sigma}_{H}\), as shown in Fig.~\ref{fig:2}(c). 

For the high-temperature regime, we evaluate the following limits separately:
\be
\label{eq:sup_F0_+}
\FD_{0}^{+}&=&\int_{1}^{\infty}dx\left(\frac{1}{e^{(x-\bar{\mu})/\bar{T}}+1}+\frac{1}{e^{(x+\bar{\mu})/\bar{T}}+1}\right)=\bar{T}\int_{1/\bar{T}}^{\infty}dx\left(\frac{1}{e^{x-\bar{\mu}/\bar{T}}+1}+\frac{1}{e^{x+\bar{\mu}/\bar{T}}+1}\right)\xrightarrow[]{T\to\infty}
2\bar{T}\ln2,
\ee
\be
\label{eq:sup_F-2_+}
\FD_{-2}^{+}&=&\int_{1}^{\infty}dx\frac{1}{x^{2}}\left(\frac{1}{e^{(x-\bar{\mu})/\bar{T}}+1}+\frac{1}{e^{(x+\bar{\mu})/\bar{T}}+1}\right)=\frac{1}{\bar{T}\rule{0pt}{2.2ex}}\int_{1/\bar{T}}^{\infty}dx\frac{d}{dx}\left(-\frac{1}{x}\right)\left(\frac{1}{e^{x-\bar{\mu}/\bar{T}}+1}+\frac{1}{e^{x+\bar{\mu}/\bar{T}}+1}\right)\nn
&=&\left[\frac{1}{e^{(1-\bar{\mu})/\bar{T}}+1}+\frac{1}{e^{(1+\bar{\mu})/\bar{T}}+1}\right]+\frac{1}{\bar{T}\rule{0pt}{2.2ex}}\int_{1/\bar{T}}^{\infty}dx\frac{d}{d x}(\ln x)\frac{d}{d x}\left(\frac{1}{e^{x-\bar{\mu}/\bar{T}}+1}+\frac{1}{e^{x+\bar{\mu}/\bar{T}}+1}\right)\nn
&=&1-\frac{1}{2\bar{T}\rule{0pt}{2.2ex}}+\mathcal{O}(\bar{T}^{-3})+\frac{1}{\bar{T}\rule{0pt}{2.2ex}}\left[\ln \bar{T}\left(-\frac{1}{2}+\mathcal{O}(\bar{T}^{-2})\right)\right]-\frac{1}{\bar{T}\rule{0pt}{2.2ex}}\underbrace{\int_{1/\bar{T}}^{\infty}dx \ln x \frac{d^{2}}{dx^{2}}\left(\frac{1}{e^{x-\bar{\mu}/\bar{T}}+1}+\frac{1}{e^{x+\bar{\mu}/\bar{T}}+1}\right)}_{\mathcal{C}=6 \log (A)-0.5 \gamma -0.5-0.66 \ln (2)\approx 0.24}\nn
&&\xrightarrow[]{T\to\infty}
1-\frac{1}{2\bar{T}\rule{0pt}{2.2ex}}\ln \bar{T}-\frac{1}{\bar{T}\rule{0pt}{2.2ex}}(\mathcal{C}+0.5),
\ee
\be
\label{eq:sup_F1_-}
\FD_{1}^{-}&=&\int_{1}^{\infty}dx\, x \left(\frac{1}{e^{(x-\bar{\mu})/\bar{T}}+1}-\frac{1}{e^{(x+\bar{\mu})/\bar{T}}+1}\right)=\bar{T}^{2}\int_{1/\bar{T}}^{\infty}dx\,x\left(\frac{1}{e^{x-\bar{\mu}/\bar{T}}+1}-\frac{1}{e^{x+\bar{\mu}/\bar{T}}+1}\right)\nn
&=&\bar{T}^{2}\frac{1}{\bar{T}\rule{0pt}{2.2ex}} \ln \left(\frac{e^{(\bar{\mu}-1)/\bar{T}}+1}{e^{-(\bar{\mu}+1)/\bar{T}}+1}\right)-\frac{1}{\bar{T}\rule{0pt}{2.2ex}}(\text{Li}_2(-e^{-(1-\bar{\mu})/\bar{T} })-\text{Li}_2(-e^{-(\bar{\mu}+1)/\bar{T}}))\xrightarrow[]{T\to\infty}2\bar{T}(\bar{\mu}\ln2),
\ee
\be
\label{eq:sup_F-3_-}
\FD_{-3}^{-}&=&\int_{1}^{\infty}dx\frac{1}{x^{3}}\left(\frac{1}{e^{(x-\bar{\mu})/\bar{T}}+1}-\frac{1}{e^{(x+\bar{\mu})/\bar{T}}+1}\right)=\frac{1}{\bar{T}^{2}\rule{0pt}{2.2ex}}\int_{1/\bar{T}}^{\infty}dx\frac{d}{dx}\left(-\frac{1}{2x^{2}}\right)\left(\frac{1}{e^{x-\bar{\mu}/\bar{T}}+1}-\frac{1}{e^{x+\bar{\mu}/\bar{T}}+1}\right)\nn
&=&-\frac{1}{2\bar{T}^{2}\rule{0pt}{2.2ex}}\left[-\frac{\bar{\mu}\bar{T}}{2}+\mathcal{O}(\bar{T}^{-1})\right]+\frac{1}{2\bar{T}^{2}\rule{0pt}{2.2ex}}\int_{1/\bar{T}}^{\infty}dx\frac{d}{d x}\left(-\frac{1}{x}\right)\frac{d}{d x}\left(\frac{1}{e^{x-\bar{\mu}/\bar{T}}+1}-\frac{1}{e^{x+\bar{\mu}/\bar{T}}+1}\right)\nn
&=&\frac{\bar{\mu}}{4\bar{T}\rule{0pt}{2.2ex}}-\frac{1}{2\bar{T}\rule{0pt}{2.2ex}}\left[\frac{\bar{\mu}}{4\bar{T}\rule{0pt}{2.2ex}}+\mathcal{O}(\bar{T}^{-3})\right]+\underbrace{\frac{1}{2\bar{T}^{2}\rule{0pt}{2.2ex}}\int_{1/\bar{T}}^{\infty}dx\frac{1}{x}\frac{d^{2}}{d x^{2}}\left(\frac{1}{e^{x-\bar{\mu}/\bar{T}}+1}-\frac{1}{e^{x+\bar{\mu}/\bar{T}}+1}\right)}_{\mathcal{O}(\bar{T}^{-3})}\xrightarrow[]{T\to\infty}\frac{\bar{\mu}}{4\bar{T}\rule{0pt}{2.2ex}},
\ee
where we used square brackets for the boundary terms, $\gamma$ for Euler's constant, $A$ for Glaisher's constant [for the constant $\mathcal{C}$ in Eq.~\eqref{eq:sup_F-2_+}], and neglected terms of order $\mathcal{O}(\bar{T}^{-2})$. The latter is because the highest temperature divergence in these terms was linear and this can only be multiplied by terms of order $\mathcal{O}(\bar{T}^{0})$ and $\mathcal{O}(\bar{T}^{-1})$ to give a nonvanishing contribution to the expansion of $\sigma_{H}^{(1)}$ as $\bar{T}\to\infty$. Also note that in this limit $\fe(\Delta)+\fh(\Delta)\to 1-1/(2\bar{T})$ and $\fe(\Delta)-\fh(\Delta)\to\bar{\mu}/(2\bar{T})$. Imposing these limits to Eq.~\eqref{eq:total_contribution_Vc}, we finally arrive at
\be
\sigma_{H}^{(1)}\xrightarrow[]{T\to\infty}\ln2(\ln \bar{T} + 2\mathcal{C}+0.5)\bar{\sigma}_{H}\approx\ln2(\ln \bar{T} +0.98)\bar{\sigma}_{H}.
\ee

\section{\label{appendix_D}Coulomb potential}

In the case of a Coulomb potential, $\mathcal{O}(k^2)$ terms no longer vanish in the limit $\Lambda\to\infty$, but instead they give a finite contribution. Since the terms containing sums and differences of Fermi--Dirac distributions remain the same as in Eqs.~\eqref{eq:sup_Sigma_0}--\eqref{eq:sup_Sigma_perpendicular}, we only look at the contribution from the Fermi sea when introducing the counterterms
\be
\label{eq:sup_Sigma_0_Coulomb}
\Sigma_0^{\text{FS}}(\bm k)-\delta\mu\ct
&=&-\frac{1}{2}\int_{\bm k\pr} V(\bm k\pr)-\delta\mu\ct=-\frac{1}{2}\int_{\bm k\pr} \frac{2\pi e^2}{\kappa k\pr(1+|\bm{k}'|^n/\Lambda^n)}-\delta\mu\ct=-\mu\underbrace{\left[\frac{\alpha}{2}\frac{\Delta}{\mu}\frac{\vf\Lambda}{\Delta\,\mbox{sinc}(\pi/n)}+\frac{\delta\mu\ct}{\mu}\right]}_{=0},
\ee
\be
\label{eq:sup_Sigma_parallel_Coulomb}
\bm \Sigma^{\text{FS}}_{\parallel}(\bm k)+\bm k \delta \vf\ct
&=&\frac{\vf}{2}\int_{\bm k\pr} V(\bm k -\bm k\pr) \frac{\bm k\pr}{\vare_{k\pr}}+\bm k \delta \vf\ct=\frac{\vf}{2}\int_{\bm k\pr} \frac{2\pi e^2}{\kappa|\bm k -\bm k\pr|\big(1+|\bm{k}-\bm{k}'|^n/\Lambda^n\big)} \frac{\bm k\pr}{\vare_{k\pr}}+\bm k \delta \vf\ct\nn
&=&\frac{\vf^2 e^2}{\kappa}\int_{0}^{1} \frac{dx}{\sqrt{x(1-x)}}\int_{\bm k\pr}\frac{\bm k\pr \big(1+|\bm{k}-\bm{k}'|^n/\Lambda^n\big)^{-1} }{x\vf^{2}(\bm k - \bm k\pr)^{2}+(1-x)(\vf^2\bm k^{\prime 2}+\Delta^{2})}+\bm k \delta \vf\ct\nn
&=&\frac{\vf^2 e^2}{\kappa}\int_{0}^{1} \frac{dx}{\sqrt{x(1-x)}}\int_{\bm k\pr}\frac{(\bm k\pr+x\bm k)\big(1+|\bm{k}'|^n/\Lambda^n\big)^{-1}}{\vf^{2}k^{\prime 2}+(1-x)(x\vf^2 k^{2}+\Delta^{2})}+\bm k \frac{e^2}{8\kappa}+\bm k \delta \vf\ct\nn
&=&\vf\bm k \underbrace{\left[\frac{\alpha}{4}\ln\left(\frac{\vf\Lambda}{\Delta}\right)+\frac{\alpha}{4}(1+\ln2)+\frac{\delta \vf\ct}{\vf}\right]}_{=0}- \vf\bm k \frac{\alpha}{4}\left[\frac{\vare_{k}-\Delta}{2(\vare_{k}+\Delta)}+\ln\left(\frac{\vare_{k}+\Delta}{2\Delta}\right)\right],
\ee
\be
\label{eq:sup_Sigma_perpendicular_Coulomb}
\Sigma_{\perp}^{\text{FS}}(\bm k)+\delta\Delta\ct
&=&\frac{\Delta}{2}\int_{\bm k\pr} V(\bm k -\bm k\pr) \frac{1}{\vare_{k\pr}}+\delta \Delta\ct=\frac{\Delta}{2}\int_{\bm k\pr} \frac{2\pi e^2}{\kappa|\bm k -\bm k\pr|\big(1+|\bm{k}-\bm{k}'|^n/\Lambda^n\big)} \frac{1}{\vare_{k\pr}}+\delta \Delta\ct\nn
&=&\frac{\Delta e^2}{\kappa} \int_{0}^{1}\frac{dx}{\sqrt{x(1-x)}}\int_{\bm k\pr}\frac{\vf\big(1+|\bm{k}-\bm{k}'|^n/\Lambda^n\big)^{-1}}{x\vf^{2}(\bm k - \bm k\pr)^{2}+(1-x)(\vf^2\bm k^{\prime 2}+\Delta^{2})}+\delta \Delta\ct\nn
&=&\frac{\Delta \vf e^2}{\kappa}\int_{0}^{1} \frac{dx}{\sqrt{x(1-x)}}\int_{\bm k\pr}\frac{\big(1+|\bm{k}'|^n/\Lambda^n\big)^{-1}}{\vf^{2}k^{\prime 2}+(1-x)(x\vf^2 k^{2}+\Delta^{2})}+\delta \Delta\ct\nn
&=&\Delta\underbrace{\left[\frac{\alpha}{2}\ln\left(\frac{\vf\Lambda}{\Delta}\right)+\frac{\alpha}{2}\ln2+\frac{\delta \Delta\ct}{\Delta}\right]}_{=0}-\frac{\alpha}{2}\Delta\ln\left(\frac{\vare_{k}+\Delta}{2\Delta}\right),
\ee
which we simplified by employing rotational symmetry. To get to the second lines of Eqs.~\eqref{eq:sup_Sigma_parallel_Coulomb} and \eqref{eq:sup_Sigma_perpendicular_Coulomb} we used the Feynman parametrization formula
\be
\frac{1}{A^{\alpha}B^{\beta}}=\frac{\Gamma(\alpha+\beta)}{\Gamma(\alpha)\Gamma(\beta)}\int_{0}^{1}dx \frac{x^{\alpha-1}(1-x)^{\beta-1}}{\big(xA+(1-x)B\big)^{\alpha+\beta}},
\ee
with terms $A=\vf^{2}(\bm k-\bm k\pr)^{2}$ and $B=\vare_{k}^{2}=\vf^{2}k^{\prime 2}+\Delta^{2}$ and exponents $\alpha=\beta=1/2$.

As discussed in the previous subsection, we choose to absorb a finite constant term into $\delta\Delta\ct$ to ensure that our definition of the energy gap corresponds to the \enquote{physical mass}. However, unlike the case of the contact potential, there is an extra $\bm k$-dependent piece in Eq.~\eqref{eq:sup_Sigma_perpendicular_Coulomb} introduced by the self-energy, but it vanishes when $\bm k = 0$.

The regularized potential used above should be read as \( V(\bm k -\bm k\pr)= \big(2\pi e^2/(\kappa|\bm{k}-\bm{k}'|)\big)\big(1+|\bm k - \bm k\pr|^n/\Lambda^{n}\big)^{-1}\), such that when changing the variable $\bm k\pr \to \bm k\pr + x\bm k$ in the third line of Eq.~\eqref{eq:sup_Sigma_parallel_Coulomb}, we are left with an extra term $\bm k e^2/(8\kappa)$. Note that there is a finite constant absorbed in the counterterm $\delta\vf\ct$ similar to the one from the mass renormalization, but this does not influence our final result, as the latter does not depend on the way the Fermi velocity is renormalized. This can be better understood by noting that $\partial_{\vf}\sigma_{H}^{(0)}=0$ back in Eq.~\eqref{eq:sup_sigma_H0}, which means that there are no counterterm diagrams introduced by the renormalization of $\vf$.

Note that in addition to the contributions from the Fermi sea given in Eqs.~\eqref{eq:sup_Sigma_0_Coulomb} and \eqref{eq:sup_Sigma_perpendicular_Coulomb}, $\Sigma_{0}$ and $\Sigma_{\perp}$ also contain temperature and chemical potential dependent pieces. The latter are integrals over Fermi--Dirac distribution functions $\fe(\vare_{k\pr})\pm\fh(\vare_{k\pr})$ and were given previously in Eqs.~\eqref{eq:sup_Sigma_0} and \eqref{eq:sup_Sigma_perpendicular}. Once we plug these together with the renormalized Fermi sea contributions given by  Eqs. \eqref{eq:sup_Sigma_0_Coulomb} and \eqref{eq:sup_Sigma_perpendicular_Coulomb} back into Eq.~\eqref{eq:all_diagrams}, we notice that the resulting integrand nontrivially depends on the angle $\varphi$ between $\bm{k}$ and $\bm{k}'$, unlike the case of the contact potential. This appears in the form $\int d\varphi V(\bm k - \bm k\pr)$ and $\int d\varphi V(\bm k - \bm k\pr)~ \bm k \cdot \bm k\pr$, which multiplied by $\kappa/(2\pi e^{2})$ can be written as
\be
\int_{0}^{2\pi}d\varphi \frac{1}{|\bm k - \bm k\pr|}&=&\int_{0}^{2\pi}d\varphi \frac{1}{\sqrt{k^{2}+k^{\prime2}-2kk\pr\cos\varphi}}=\int_{0}^{2\pi}d\varphi \frac{1}{\sqrt{k^{2}+k^{\prime2}-2kk\pr\cos(\varphi+\pi)}}\nn
&=&2\int_{0}^{\pi}d\varphi \frac{1}{\sqrt{k^{2}+k^{\prime2}+2kk\pr\cos\varphi}}=2\int_{0}^{\pi}d\varphi \frac{1}{\sqrt{(k+k\pr)^2-4kk\pr\sin^{2}(\varphi/2)}}\nn
&=&\frac{4}{k+k\pr}\int_{0}^{\pi/2}d\varphi \frac{1}{\sqrt{1-\frac{4kk\pr}{(k+k\pr)^{2}}\sin^{2}\varphi}}=\frac{4}{k+k\pr}K\left(\frac{2\sqrt{kk\pr}}{k+k\pr}\right),
\ee
\be
\int_{0}^{2\pi} d\varphi \frac{\bm k \cdot \bm k\pr}{|\bm k - \bm k\pr|}&=&2\int_{0}^{\pi} d\varphi \frac{kk\pr\cos\varphi}{\sqrt{k^2+k^{\prime2}-2kk\pr\cos\varphi}}=-2\int_{0}^{\pi} d\varphi \frac{kk\pr\Big(1-2\sin^2(\varphi/2)\Big)}{\sqrt{k^2+k^{\prime2}+2kk\pr\big(1-2\sin^2(\varphi/2)\big)}}\nn
&=&-4kk\pr\int_{0}^{\pi/2} d\varphi \frac{1}{\sqrt{(k+k^{\prime})^2-4kk\pr\sin^2\varphi}}+8kk\pr\int_{0}^{\pi/2} d\varphi \frac{\sin^2\varphi}{\sqrt{(k+k^{\prime})^2-4kk\pr\sin^2\varphi}}\nn
&=&-\frac{4kk\pr}{k+k\pr}K\left(\frac{2\sqrt{kk\pr}}{k+k\pr}\right)+2(k+k\pr)\int_{0}^{\pi/2}d\varphi \frac{\frac{4kk\pr}{(k+k\pr)^{2}}\sin^{2}\varphi}{\sqrt{1-\frac{4kk\pr}{(k+k\pr)^{2}}\sin^{2}\varphi}}\nn
&=&-\frac{4kk\pr}{k+k\pr}K\left(\frac{2\sqrt{kk\pr}}{k+k\pr}\right)+2(k+k\pr)\left(-E\left(\frac{2\sqrt{kk\pr}}{k+k\pr}\right)+K\left(\frac{2\sqrt{kk\pr}}{k+k\pr}\right)\right)\nn
&=&\frac{2(k^2+k^{\prime2})}{k+k\pr}K\left(\frac{2\sqrt{kk\pr}}{k+k\pr}\right)-2(k+k\pr)E\left(\frac{2\sqrt{kk\pr}}{k+k\pr}\right),
\ee
where $K$ and $E$ are complete elliptic integrals of the first and second kind, respectively~\cite{gradshteyn2014table}. Plugging these integrals into Eq.~\eqref{eq:all_diagrams}, we find the Hall conductivity
\begin{align}
\sigma_{H}^{(1)}=&-\frac{e^{2}}{2h}\alpha\frac{1}{4\pi}\int_{0}^{\infty} d\bar{k}\,\bar{k}\int_{0}^{\infty} d\bar{k}\pr\,\bar{k}\pr \Bigg\{K\left(\frac{2\sqrt{\bar{k}\bar{k}\pr\rule{0pt}{2.2ex}}}{\bar{k}+\bar{k}\pr\rule{0pt}{2.2ex}}\right)\Bigg[\frac{4}{\bar{k}+\bar{k}\pr\rule{0pt}{2.2ex}}\left(\frac{1}{\bar{\vare}_{k}^{3}}\frac{\partial \Phi^{+}_{k}}{\partial\bar{\vare}_{k}}(\Phi^{+}_{k\pr}-1)+\frac{1}{\bar{\vare}_{k}^{2}\bar{\vare}_{k\pr}}\frac{\partial \Phi^{-}_{k}}{\partial\bar{\vare}_{k}}(\Phi^{-}_{k\pr}-1)\right)\nn
&-\frac{2(\bar{k}-\bar{k}\pr)}{\bar{\vare}_{k}^{2}\bar{\vare}_{k\pr}^{3}}\frac{\partial\Phi^{-}_{k}}{\partial\bar{\vare}_{k}}\Phi^{-}_{k\pr}\Bigg]-E\left(\frac{2\sqrt{\bar{k}\bar{k}\pr\rule{0pt}{2.2ex}}}{\bar{k}+\bar{k}\pr\rule{0pt}{2.2ex}}\right)\frac{2(\bar{k}+\bar{k}\pr)}{\bar{\vare}_{k}^{2}\bar{\vare}_{k\pr}^{3}}\frac{\partial \Phi^{-}_{k}}{\partial\bar{\vare}_{k}}\Phi^{-}_{k\pr}\Bigg\}+\frac{e^{2}}{2h}\frac{\alpha}{2}\int_{0}^{\infty} d\bar{k} \frac{\bar{k}}{\bar{\vare}_{k}^{2}}\frac{\partial \Phi^{-}_{k}}{\partial\bar{\vare}_{k}}\ln\left(\frac{\bar{\vare}_{k}+1}{2}\right),
\end{align}
where all the barred quantities are dimensionless $\bar{k}=\vf k/\Delta$, $\bar{\vare}_{k}=\vare_{k}/\Delta$. This expression can be evaluated numerically as a function of both temperature and chemical potential, and its contribution added to $\sigma_{H}^{(0)}$ is shown in Figs.~\ref{fig:3}(c)--\ref{fig:3}(f) for different values of $\alpha$.

\twocolumngrid

\end{document}